\documentclass[12pt]{article}

\usepackage{latexsym,epsf,cite,epsfig,amssymb}

\def\spose#1{\hbox to 0pt{#1\hss}}
\def\ltapprox{\mathrel{\spose{\lower 3pt\hbox{$\mathchar"218$}}
 \raise 2.0pt\hbox{$\mathchar"13C$}}}
\def\gtapprox{\mathrel{\spose{\lower 3pt\hbox{$\mathchar"218$}}
 \raise 2.0pt\hbox{$\mathchar"13E$}}}

\def\case#1#2{{\textstyle\frac{#1}{#2}}}

\begin{document}
\begin{titlepage}
\begin{center}
{\Large\bf Field-theory results for three-dimensional transitions with complex symmetries}
\end{center}
\vskip 1.3cm
\centerline{
Pasquale Calabrese,$^a$ 
Andrea Pelissetto,$^b$ 
Paolo Rossi,$^c$ 
Ettore Vicari$^{c\;*}$}

\vskip 0.4cm
\centerline{\sl  $^a$ Scuola Normale Superiore, Pisa, Italy} 
\centerline{\sl  $^b$ Dipartimento di Fisica dell'Universit\`a di Roma I,
                       Roma, Italy}
\centerline{\sl  $^c$ Dipartimento di Fisica dell'Universit\`a di Pisa,
                       Pisa, Italy}

\vskip 0.4cm
\begin{center}
E-mail: {\tt calabres@df.unipi.it, Andrea.Pelissetto@roma1.infn.it,} \\

{\tt Paolo.Rossi@df.unipi.it, Ettore.Vicari@df.unipi.it}
\end{center}

\vskip 1.cm

\begin{abstract}
We discuss several examples of three-dimensional critical phenomena
that can be described by Landau-Ginzburg-Wilson $\phi^4$ theories.
We present an overview of field-theoretical results
obtained from the analysis of high-order perturbative
series in the frameworks of the $\epsilon$ and of the fixed-dimension $d=3$
expansions.  In particular, we discuss 
the stability of the O($N$)-symmetric fixed
point in a generic $N$-component theory,
the critical behaviors of randomly dilute Ising-like systems
and frustrated spin systems with noncollinear order,
the multicritical behavior arising from the competition
of two distinct types of ordering with symmetry O($n_1$) and O($n_2$) 
respectively. 

\end{abstract}

\vskip 5.cm

\noindent
$*$
Talk given at the International Conference of Theoretical Physics, TH2002, 
Paris, July 22-27, 2002.

\end{titlepage}

\section{Introduction}
\label{lsec-intro}

In the framework of the renormalization-group (RG)
approach to critical phenomena,
a quantitative description of 
many continuous phase transitions  can be obtained
by considering an effective Landau-Ginzburg-Wilson (LGW) theory,
containing up to fourth-order powers of the field components.
The simplest example is the O($N$)-symmetric $\phi^4$ theory,
\begin{equation}
{\cal H}_{O(N)} = \int d^d x \Bigl[ 
{1\over 2} \sum_i (\partial_\mu \Phi_{i})^2 + 
{1\over 2} r \sum_i \Phi_{i}^2  + 
{1\over 4!} u \sum_{ij} \Phi_i^2 \Phi_j^2 \Bigr], 
\label{HON}
\end{equation}
where $\Phi$ is an $N$-component real field. This model
describes several universality classes:
the Ising one for $N=1$ 
(e.g., liquid-vapor transition), the XY one for $N=2$
(e.g., superfluid transition in $^4$He),
the Heisenberg one for $N=3$
(isotropic magnets), and long self-avoiding
walks for $N\rightarrow 0$
(dilute polymers).
See, e.g., Refs.~\cite{review,Zinn-Justin-book} for 
recent reviews.
But there are  also other physically interesting 
transitions described by LGW theories
characterized by more complex symmetries.

The general LGW Hamiltonian for an $N$-component field $\Phi_i$
can be written as
\begin{equation}
{\cal H} = \int d^d x \Bigl[ 
{1\over 2} \sum_i (\partial_\mu \Phi_{i})^2 + 
{1\over 2} \sum_i r_i \Phi_{i}^2  + 
{1\over 4!} \sum_{ijkl} u_{ijkl} \; \Phi_i\Phi_j\Phi_k\Phi_l \Bigr], 
\label{generalH}
\end{equation}
where the number of independent parameters $r_i$ and $u_{ijkl}$ 
depends on the symmetry group of the theory. An interesting 
class of models are those in which $\sum_i \Phi^2_i$ is the 
only quadratic invariant polynomial.
In this case, all $r_i$ are equal, $r_i = r$, and 
$u_{ijkl}$ satisfies the trace condition~\cite{BLZ-76}
$\sum_i u_{iikl} \propto \delta_{kl}$.
In these models, criticality is driven by tuning the single parameter
$r$. Therefore, they describe critical phenomena characterized 
by one (parity-symmetric) relevant parameter, 
which often corresponds to the temperature. Of course, there is also
(at least one) parity-odd relevant parameter that corresponds to 
a term $\sum_i h_i \Phi_i$ that can be added to the Hamiltonian
(\ref{generalH}). For symmetry reasons, criticality is observed 
for $h_i\to 0$. 
There are several physical systems whose critical behavior 
can be described by this type of LGW Hamiltonians with two or more quartic 
couplings, see, e.g., Refs.~\cite{Aharony-76,review}.
More general LGW Hamiltonians, that allow for the presence
of independent quadratic parameters $r_i$, must be considered
to describe multicritical behaviors 
arising from the competition of distinct types of ordering.
A multicritical point can be observed at the intersection of 
two critical lines with different order parameters.
In this case the multicritical behavior is achieved
by tuning two relevant scaling fields, which may correspond to
the temperature and to an anisotropy parameter \cite{KNF-76}.

In the field-theory (FT) approach the RG flow 
is determined by a set of RG
equations for the correlation functions of the order parameter.
In the case of a continuous transition, 
the critical behavior is determined by the stable fixed point (FP)
of the theory, which characterizes a universality class. 
The absence of a stable FP is instead 
an indication for a first-order transition,
even in those cases in which the mean-field approximation predicts
a continuous transition.
But, even in the presence of a stable FP,
a first-order transition may  still occur
for systems that are outside its attraction domain. 
The RG flow can be studied by perturbative methods, by performing an 
expansion in powers of $\epsilon\equiv 4-d$\cite{WF-72} or a 
fixed-dimension (FD) expansion in powers of appropriate zero-momentum
quartic couplings \cite{Parisi-80}. 
In these perturbative approaches,
the computation and the resummation of high-order series
is essential to obtain reliable results for three-dimensional transitions
(see Refs.~\cite{Zinn-Justin-book,review} for reviews). 
Beside improving the accuracy, in some cases high-order
calculations turn out to be necessary to determine 
the correct physical picture in three dimensions.

In this paper we give an overview of the perturbative FT
results obtained for a number of three-dimensional transitions 
described by LGW Hamiltonians.
In Sec.~\ref{onstab} we discuss the stability of the 
O($N$)-symmetric fixed point under generic perturbations in  
three-dimensional $N$-component systems.
In Sec.~\ref{random} we discuss the critical behavior
of Ising-like systems with quenched disorder
effectively coupled to the energy, for instance the 
randomly dilute Ising model.
In Sec.~\ref{chiral} we consider frustrated spin models
with noncollinear order, whose critical behavior is effectively described
by an O($M$)$\otimes$O($N$)-symmetric Hamiltonian with $M=2$.
Finally, in Sec.~\ref{multi} we discuss the predictions of 
the O($n_1$)$\oplus$O($n_2$)-symmetric $\phi^4$ theory for 
the multicritical behavior observed near the point
where two critical lines with symmetry O($n_1$) and O($n_2$) meet.

\section{Stability of the O($N$)-symmetric fixed point}
\label{onstab}

In order to discuss the stability of the O($N$) FP  in a 
generic $N$-component system, 
it is convenient to consider polynomial perturbations  
$P_{m,l }^{a_1,\ldots,a_{l} }$, 
$m\ge l$, which are of degree $m$ in the $N$-component field $\Phi^a$
and transform as the $l$-spin representation of the O($N$) group.
Explicitly formulae can be found in Ref.~\cite{CPV-02-mc}.
In addition, one should also consider perturbations containing
derivatives of the field. 
At least near four dimensions, one can use
standard RG arguments to show that, 
beside the O($N$)-symmetric terms $P_{2,0}=\Phi^2$ and 
$P_{4,0}=(\Phi^2)^2$, only three other perturbations should be 
considered, $P_{2,2}^{ab}$, $P_{4,2}^{ab}$, and 
$P_{4,4}^{abcd}$.
The stability properties 
of the O($N$) FP depend on the RG dimensions $y_{m,l}$ of these 
perturbations.\footnote{Note that $P_{2,2}^{ab}$ and 
$P_{4,4}^{abcd}$ are RG eigenoperators, while $P_{4,2}^{ab}$ 
must be in general properly subtracted, i.e. the 
RG eigenoperator is $P_{4,2}^{ab} + z P_{2,2}^{ab}$ for a suitable value of $z$.}

In Table~\ref{ONstab} we report FT estimates of 
the RG dimensions $y_{m,l}$
for $N=2,3,4,5$, obtained from the analysis of six-loop FD
and five-loop $\epsilon$ series 
\cite{CPV-02,CPV-02-mc,GZ-98}.\footnote{Results obtained in 
other theoretical approaches and in experiments  
can be found in Refs.~\cite{CPV-02-mc,review} and references therein.}
The quadratic perturbations $P_{2,2}^{ab}$ are relevant for the description of 
the breaking of the O($N$) symmetry down to O($M$)$\oplus$O($N-M$).
Since $y_{2,2}>0$, they are always relevant.
The RG dimension $y_{4,2}$ is negative for any $N$, so that the corresponding
spin-2 perturbation $P_{4,2}^{ab}$ is always irrelevant.
On the other hand, the sign of $y_{4,4}$ depends on $N$:
it is clearly negative for $N=2$ and positive for $N\ge 4$.
For $N=3$ it is marginally positive, suggesting the instability
of the O(3) FP under generic spin-4 quartic perturbations.
Actually the stability of the O($N$) FP 
can be inferred from the RG flow of the cubic-symmetric LGW Hamiltonian
for an $N$-component field
\begin{equation}
{\cal H}_{c} =  \int d^d x\, \left\{ {1\over 2} \sum_{i=1}^{N}
      \Bigl[ (\partial_\mu \Phi_i)^2 +  r \Phi_i^2 \Bigr]  
+{1\over 4!} \Bigl[ u (\sum_i^N \Phi_i^2)^2 + v \sum_i^N \Phi_i^4 \Bigr]
\right \} .
\label{Hphi4cubic}
\end{equation}
The point is that the cubic-symmetric 
perturbation $\sum_i \Phi_i^4$
is a particular combination of the spin-4 operators $P_{4,4}^{abcd}$ and 
of the spin-0 term $P_{4,0}$.  
The RG flow for the cubic-symmetric theory has been much investigated
using various FT and lattice techniques \cite{review}. 
The O($N$) FP turns out to be unstable for $N>N_c$ with $N_c\approx 3$.
The most accurate results have been provided by analyses of 
high-order FT perturbative expansions, six-loop FD 
and five-loop $\epsilon$ series, see e.g. Refs.~\cite{CPV-00,FHY-00},
which find $N_c\ltapprox 2.9$ in three dimensions, and the existence of 
a stable FP characterized by a reduced cubic symmetry for $N\geq N_c$.
These results imply that
the O($N$)-symmetric FP is unstable under spin-4 quartic perturbations 
for $N\ge 3$,
and can be applied to establish the stability of the O($N$) FP
in any physical critical phenomenon that is effectively described
by a generic LGW Hamiltonian for an $N$-component 
field.\footnote{
Note that the condition that $\sum \Phi^2_i$ is the only quadratic 
invariant forbids the presence in the Hamiltonian 
of any spin-2 term $P^{ab}_{2,2}$. Analogously,
the trace condition $\sum_i u_{iikl} \propto \delta_{kl}$
forbids quartic polynomials transforming as the spin-2 
representation of the O($N$) group, i.e. the operators $P_{4,2}^{ab}$.}

\begin{table*}
\caption{
Three-dimensional estimates of the RG dimensions $y_{m,l}$
from $\epsilon$ and  FD expansions.
}
\label{ONstab}
\footnotesize
\begin{center}
\begin{tabular}{lclllllll}
\hline
\multicolumn{1}{c}{$N$}&
\multicolumn{1}{c}{}&
\multicolumn{1}{c}{$y_{2,0}=\nu^{-1}$}&
\multicolumn{1}{c}{$y_{2,2}$}&
\multicolumn{1}{c}{$\phi_T\equiv y_{2,2}\nu$}&
\multicolumn{1}{c}{$y_{4,0}$}&
\multicolumn{1}{c}{$y_{4,2}$}&
\multicolumn{1}{c}{$y_{4,4}$}&
\multicolumn{1}{c}{$\phi_{4,4}\equiv y_{4,4}\nu$}\\
\hline
2 &$\epsilon$ & 1.497(8) & 1.766(6) & 1.174(12)    & $-$0.802(18) & $-$0.624(10) &$-$0.114(4) &$-$0.077(3)\\
  & FD        & 1.493(3) &          & 1.184(12)    & $-$0.789(11) &              & $-$0.103(8)&$-$0.069(5) \\ \hline

3 &$\epsilon$ & 1.419(11)& 1.790(3) & 1.260(11)& $-$0.794(18) & $-$0.550(14) & 0.003(4)& 0.002(3) \\ 
  & FD        & 1.414(7) &          & 1.27(2)  & $-$0.782(13) & &  0.013(6)& 0.009(4) \\ \hline

4&$\epsilon$  & 1.357(15)& 1.813(6) & 1.329(16)& $-$0.795(30) & $-$0.493(14) & 0.105(6)& 0.079(5) \\ 
& FD          & 1.350(11)&          & 1.35(4) & $-$0.774(20) & & 0.111(4)& 0.083(3)  \\ \hline

5  &$\epsilon$ & 1.333(36) & 1.832(8) & 1.40(3) & $-$0.783(26) & $-$0.441(13) & 0.198(11) &  0.151(9) \\ 
& FD & 1.312(12) & & 1.40(4) & $-$0.790(15) & &  0.189(10)& 0.144(8)  \\ \hline

$\infty$ & & 1 & 2 & 2 & $-$1 & 0 & 1 & 1 \\
\hline
\end{tabular}
\end{center}
\end{table*}

\section{Randomly dilute Ising model}
\label{random}

In the last few decades many theoretical and experimental
studies have investigated the critical properties
of statistical models in the presence of quenched disorder.
A typical example is obtained by mixing a
uniaxial antiferromagnet with a nonmagnetic material,
such as Fe$_x$Zn$_{1-x}$F$_2$ and Mn$_x$Zn$_{1-x}$F$_2$.
These materials can be modeled by the randomly dilute Ising model (RDIM)
\begin{equation}
{\cal H}_{\rm RDIM} = J\,\sum_{<ij>}  \rho_i \,\rho_j \; s_i s_j,
\label{latticeH}
\end{equation}
where the sum is extended over all nearest-neighbor sites of a 
lattice,
$s_i=\pm 1$ are the spin variables,
$\rho_i$ are uncorrelated quenched random variables, which are equal to one 
with probability $x$ (the spin concentration) and zero with probability $1-x$
(the impurity concentration). 
Above the percolation threshold of the spin concentration,
the critical behavior of the RDIM belongs to a new 
universality class that is distinct from the Ising universality class
of pure systems, and that is shared by all systems with quenched
disorder effectively coupled to the energy.
See, e.g., Refs. \cite{Belanger-00,FHY-01,review} for recent reviews.

Using the FT approach and the replica method, one arrives at
the effective LGW Hamiltonian ${\cal H}_c$ \cite{GL-76}, 
cf. Eq.~(\ref{Hphi4cubic}),
which is expected to describe the critical properties of the RDIM
in the limit $N\rightarrow 0$.
The most precise FT results for the critical exponents
have been obtained by analyzing the FD six-loop expansions \cite{PV-00,PS-00}.
The major drawback of the FT perturbative approach is the non-Borel summability 
of the series due to a more complicated analytic structure of the field theory  
corresponding to quenched disordered models. 
Nevertheless, series analyses seem to provide 
sufficiently robust estimates, which are in good agreement 
with experiments and recent Monte Carlo simulations.
The results of the six-loop analysis 
are reported in Table~\ref{rimexp}, where they are 
compared with estimates obtained in Monte Carlo simulations of the RDIM 
and in experiments on uniaxial magnets.
The values of the exponents are definetely different
from those of the pure Ising universality class, where, e.g.,
$\nu=0.63012(16)$ \cite{CPRV-ising}.

\begin{table*}
\caption{
Critical exponents for the RDIM universality class.}
\label{rimexp}
\footnotesize
\begin{center}
\begin{tabular}{ccllll}
\hline
\multicolumn{1}{c}{}& 
\multicolumn{1}{c}{}& 
\multicolumn{1}{c}{$\gamma$}& 
\multicolumn{1}{c}{$\nu$}& 
\multicolumn{1}{c}{$\alpha$}&
\multicolumn{1}{c}{$\beta$}\\ 
\hline  
six-loop FD & \cite{PV-00} & 1.330(17) & 0.678(10)  & $-$0.034(30) & 0.349(5)\\
Monte Carlo & \cite{BFMMPR-98}  & 1.342(10) & 0.684(5)   & $-$0.051(16) & 0.3546(28) \\
Fe$_x$Zn$_{1-x}$F$_2$  & \cite{Belanger-00}  & 1.31(3) & 0.69(1) & $-$0.10(2) & 0.359(9) \\
\hline
\end{tabular}
\end{center}
\end{table*}

Using the FT approach, one can also compute the critical exponent $\phi$ 
describing the crossover from random-dilution to random-field
critical behavior in Ising systems,
and in particular the crossover observed in dilute anisotropic
antiferromagnets when an external magnetic field is applied \cite{Belanger-00}.
The crossover exponent $\phi$ is related to the RG dimensions of
the quadratic operator $\Phi_i\Phi_j$ ($i\neq j$) in the 
limit $N\rightarrow 0$ \cite{Aharony-86}.
Six-loop computations \cite{CPV-inprep}
provide the estimate $\phi=1.43(1)$,
which turns out to be in
good agreement with the available experimental estimates,
for example $\phi=1.42(3)$ for Fe$_x$Zn$_{1-x}$F$_2$ \cite{Belanger-00}.

Finally, we mention that six-loop perturbative series for multicomponent
systems with quenched disorder, taking also into account a possible
cubic anisotropy, have been computed and analyzed in Refs.~\cite{PV-00,CPV-03}.

\section{Frustrated spin models with noncollinear order}
\label{chiral}

In physical magnets noncollinear order is due to frustration that may arise
either because of the special geometry of the lattice, or from the competition 
of different kinds of interactions \cite{Kawamura-98}.
Typical examples of systems of the first type are 
stacked triangular antiferromagnets (STA's), 
where magnetic ions are located at each site of 
a three-dimensional stacked triangular lattice.
On the basis of the structure of the ground state, 
in an $N$-component STA one expects a 
transition associated with a breakdown of 
the symmetry from O($N$) in the HT phase to O($N-2$) in the LT phase. 
The nature of the transition is still controversial. In 
particular, the question is whether the critical behavior belongs
to a new chiral universality class, as originally conjectured by Kawamura \cite{Kawamura-88}.  
On this issue, there is still much 
debate, FT methods, Monte Carlo simulations, and experiments providing 
contradictory results in many cases (see, e.g., Ref.\cite{review}
for a recent review of results).
Overall, experiments on STA's favor a continuous transition
belonging to a new chiral universality class.

The determination of an effective LGW Hamiltonian describing
the critical behavior leads to
the O($M$)$\otimes$O($N$)-symmetric theory \cite{Kawamura-88}
\begin{equation}
{\cal H}_{ch}  = \int d^d x
 \Bigl\{ {1\over2}
      \sum_{a} \Bigl[ (\partial_\mu \phi_{a})^2 + r \phi_{a}^2 \Bigr]
+ {1\over 4!}u \Bigl( \sum_a \phi_a^2\Bigr)^2
+ {1\over 4!}  v
\sum_{a,b} \Bigl[ ( \phi_a \cdot \phi_b)^2 - \phi_a^2\phi_b^2\Bigr]
             \Bigr\},
\label{LGWH}
\end{equation}
where $\phi_a$, $a=1,\ldots, M$, are $N$-component vectors. 
The case $M=2$ with $v>0$ describes  
frustrated spin models with noncollinear order;
\footnote{Negative values of $v$ correspond to magnets 
with sinusoidal spin structures.}
$N=2$ and $N=3$ correspond to
XY and Heisenberg systems, respectively.  
Recently the Hamiltonian (\ref{LGWH}) has been also considered
to discuss the phase diagram of Mott insulators \cite{Sachdev-02}.
See Refs.~\cite{Kawamura-98,review} for other applications.

Six-loop calculations \cite{PRV-01} in the framework of the
$d=3$ FD expansion provide a rather robust evidence
for the existence of a new stable FP in the
XY and Heisenberg cases corresponding to the conjectured
chiral universality class, and  contradicting earlier studies
based on much shorter (three-loop) series \cite{AS-94}.
It has also been argued \cite{CPS-02} that the stable chiral FP
is actually a focus, due to the fact that the eigenvalues of 
its stability matrix turn out to have a nonzero imaginary part.
The new chiral FP's found for $N=2,3$
should describe the apparently continuous transitions observed in STA's.
The FT estimates of the critical exponents are in satisfactory agreement with
the experimental results, including the chiral crossover exponent
related to the chiral degrees of freedom \cite{PRV-02}. 
We also mention that high-order FT analyses of 
two-dimensional systems have been reported in Ref.~\cite{CP-01}.

On the other hand, other FT studies, see, e.g., Ref. \cite{TDM-00},
based on approximate solutions of continuous RG equations,
do not find a stable FP, thus favoring a weak first-order transition.
Monte Carlo simulations have not been conclusive in setting the question, 
see, e.g., Refs.~\cite{LS-98,Itakura-01,PS-02}.
Since all the above approaches rely on  
different approximations and assumptions, their comparison and consistency 
is essential before considering the issue substantially understood.

\section{Multicritical behavior in O($n_1$)$\oplus$O($n_2$) theories}
\label{multi}

The competition of distinct types of ordering gives rise to multicritical
behavior. More specifically, a multicritical point (MCP) is observed at the 
intersection of two critical lines characterized by different order parameters.
MCP's arise in several physical contexts, for instance 
in anisotropic antiferromagnets, 
in high-$T_c$ superconductors, in $^4$He, etc.
The multicritical behavior arising from the
competition of two orderings characterized by O($n$) 
symmetries is determined by the RG flow of  the most general
O($n_1$)$\oplus$O($n_2$)-symmetric
LGW Hamiltonian involving two fields $\phi_1$ and $\phi_2$
with $n_1$ and $n_2$ components respectively, i.e.
\cite{KNF-76} 
\begin{eqnarray}
{\cal H}_{mc} = \int d^d x \Bigl[ 
\case{1}{2} ( \partial_\mu \phi_1)^2  + \case{1}{2} (
\partial_\mu \phi_2)^2 + \case{1}{2} r_1 \phi_1^2  
 + \case{1}{2} r_2 \phi_2^2  
+ u_1 (\phi_1^2)^2 + u_2 (\phi_2^2)^2 + w \phi_1^2\phi_2^2 \Bigr].
\label{bicrHH} 
\end{eqnarray}
The critical behavior at the MCP is determined 
by the stable FP when both $r_1$ and $r_2$ 
are tuned to their critical value.
An interesting possibility is that the stable FP has O($N$) symmetry, 
$N\equiv n_1 + n_2$, so that the symmetry gets effectively enlarged  
approaching the MCP.

The phase diagram of the model with
Hamiltonian (\ref{bicrHH}) has been investigated
within the mean-field approximation in Ref. \cite{LF-72}.
If the transition at the MCP is continuous, one may observe 
either a bicritical or a tetracritical behavior.
But it is also possible that the transition at the MCP is of first order.
$O(\epsilon)$  calculations in the framework of the 
$\epsilon$ expansion \cite{KNF-76} 
show that the isotropic O($N$)-symmetric FP ($N\equiv n_1 + n_2$)
is stable for $N<N_c=4 + O(\epsilon)$.
With increasing $N$, a new FP named biconal FP (BFP),
which has only O($n_1$)$\oplus$O($n_2$) symmetry, becomes stable. 
Finally, for large $N$, the decoupled FP (DFP)
is the stable one. In this case, the two order parameters are 
effectively uncoupled at the MCP, giving rise to a tetracritical 
behavior. 

The $O(\epsilon)$ computations provide useful indications on the 
RG flow in three  dimensions, but a controlled 
extrapolation to $\epsilon=1$ requires much longer series
and an accurate resummation exploiting their Borel summability.
For this purpose we have extended the $\epsilon$ expansion
to $O(\epsilon^5)$ \cite{CPV-02-mc}.  
A robust picture of the RG flow predicted by
the  O($n_1$)$\oplus$O($n_2$)-symmetric
LGW theory can be achieved 
by supplementing the analysis of the $\epsilon$ series 
with the results for the stability of
the O($N$) FP (cf. Sec.~\ref{onstab}),
which were also obtained by analyzing six-loop FD series,
and with nonperturbative arguments allowing to establish
the stability of the DFP \cite{Aharony-02}.
Since the Hamiltonian (\ref{bicrHH}) contains spin-4 quartic perturbations
with respect to the O($N$) FP,
the results for the spin-4 RG dimension $y_{4,4}$ 
(cf. Table~\ref{ONstab})
imply that the O($N$) FP is stable only for $N=2$, i.e.
when two Ising-like critical lines meet.
It is  unstable in all cases with $N\ge 3$.
This implies that for $N\ge 3$ the enlargement of the symmetry 
O($n_1$)$\oplus$O($n_2$) to O($N$) does not occur, unless
an additional parameter is tuned
beside those associated with the quadratic perturbations. 
For $N=3$, i.e. for $n_1=1$ and $n_2=2$,
 the critical behavior at the MCP is described by the BFP,
whose critical exponents turn out to be very close to those of
the Heisenberg universality class.
For $N\ge 4$ and for any $n_1,n_2$
the DFP is stable, implying a tetracritical behavior.
This can also be inferred by using nonperturbative arguments 
\cite{Aharony-02}
that allow to determine the relevant stability eigenvalue
from the critical exponents of the O($n_i$)
universality classes.

Anisotropic antiferromagnets in a uniform magnetic field $H_\parallel$
parallel to the anisotropy axis present a MCP 
in the $T-H_\parallel$ phase diagram, where two critical lines
belonging to the XY and Ising universality classes meet \cite{KNF-76}. 
The above results predict a multicritical behavior
described by the BFP, contradicting the $O(\epsilon)$ calculations
that suggested the stability of the O(3) FP.
Notice that it is hard to distinguish 
the biconal from the  O(3) critical behavior.
For instance, the correlation-length exponent
$\nu$ differs by less than 0.001 in the two cases.

The case $N=5$, $n_1=2$, $n_2=3$ is relevant for the 
SO(5) theory \cite{Zhang-97,ZHAHA-99} of high-$T_c$ superconductors,
which proposes a description of these materials in terms of a three-component
antiferromagnetic order parameter and 
a $d$-wave superconducting order parameter with U(1) symmetry,
with an approximate O(5) symmetry.
Within the SO(5) theory, it has been speculated that 
the antiferromagnetic and superconducting transition lines meet
at a MCP in the temperature-doping phase diagram,
which is bicritical and shows an effectively enlarged O(5) symmetry.
There are also recent claims in favor of the stability of the
O(5) FP based on Monte Carlo simulations
of three-dimensional five-component systems \cite{Hu-01}.
Our results on the RG flow of the O(2)$\oplus$O(3) theory
show that the O(5) FP cannot describe the asymptotic critical behavior
at the MCP, unless a further tuning of the parameters is performed.
Therefore, the O(5) symmetry is not effectively realized at the 
point where the antiferromagnetic and superconducting transition 
lines meet. The multicritical behavior is either governed by the tetracritical
decoupled fixed point or is of first-order type if the system is outside 
its attraction domain.
The predicted tetracritical behavior may explain a number of recent experiments
that provided evidence of a coexistence region
of the antiferromagnetic and superconducting phases, see, e.g., 
Ref.~\cite{ZDS-02}.
The O(5) FP is unstable with a crossover exponent $\phi_{4,4}\approx 0.15$,
which, although rather small, is nonetheless sufficiently large  
not to exclude the possibility of observing 
the RG flow towards the eventual asymptotic behavior 
for reasonable values of the reduced temperature,
even in systems with a moderately small breaking of the O(5) symmetry,
such as those described by the projected SO(5) model discussed in 
Refs.~\cite{ZHAHA-99,Dorneich-etal-02}.


\end{document}